\documentclass[aps,prl,twocolumn,groupedaddress]{revtex4}
\usepackage{amssymb,graphicx,xspace,color}
\begin{document}

\title{Probing the Unconventional Superconducting State of LiFeAs by Quasiparticle Interference}
\author
{Torben H\"{a}nke, Steffen Sykora, Ronny Schlegel, Danny Baumann, Luminita Harnagea, Sabine Wurmehl, Maria Daghofer, Bernd B\"{u}chner, Jeroen van den Brink}
\affiliation{IFW Dresden, 01171 Dresden, Germany}
\author
{Christian Hess}
\email[]{c.hess@ifw-dresden.de}
\affiliation{IFW Dresden, 01171 Dresden, Germany}

\date{\today}

\begin{abstract}
A crucial step in revealing the nature of unconventional superconductivity is to investigate the symmetry of the superconducting order parameter. Scanning tunneling spectroscopy has proven a powerful technique to probe this symmetry by measuring the quasiparticle interference (QPI) which sensitively depends on the superconducting pairing mechanism.  A particularly well suited  material to apply this technique is the stoichiometric superconductor LiFeAs as it features clean, charge neutral cleaved surfaces without surface states and a relatively high $T_c\sim18$~K. Our data reveal that in LiFeAs the quasiparticle scattering is governed by a van-Hove singularity at the center of the Brillouin zone which is in stark contrast with other pnictide superconductors where nesting is crucial for both scattering and $s_\pm$-superconductivity. 
Indeed, within a minimal model and using the most elementary order parameters, calculations of the QPI suggest a dominating role of the hole-like bands for the quasiparticle scattering. Our theoretical findings do not support the elementary singlet pairing symmetries $s_{++}$, $s_\pm$, and $d$-wave. 
This brings to mind that the superconducting pairing mechanism in LiFeAs is based on an unusual pairing symmetry such as an elementary $p$-wave (which  provides optimal agreement between the experimental data and QPI simulations) or a more complex order parameter (e.g. $s+id$-wave symmetry).
\end{abstract}


\maketitle 
The discovery of iron-based superconductors \cite{Kamihara2008} has generated enormous research activities to reveal the nature of superconductivity in these materials. 
In particular, $s_\pm$-pairing, i.e. an $s$-wave order parameter with alternating sign between almost perfectly nested hole and electron pockets has been suggested to be prevailing for the entire class of iron-based superconductors \cite{Mazin2008,Kuroki2008,Mazin2010}. In this regard, the material LiFeAs is of particular interest since experiments have proven an absence of nesting \cite{Borisenko2010} and theoretical works yield contradictory results, i.e., both $s_\pm$-wave singlet as well as $p$-wave triplet pairing has been suggested \cite{Brydon2011,Thomale2011}. 

A powerful method to probe the symmetry of the superconducting order parameter is to map out the spatial dependence of the local density of states (DOS) by scanning tunneling spectroscopy (STS). In such experiments, the relation $\mathrm{DOS}(E)\propto dI/dV(V_{\mathrm{bias}})$ with tunneling voltage $V$ and current $I$ at energy $E$ and bias voltage $V_{\mathrm{bias}}$ ($E=eV_{\mathrm{bias}}$) is exploited. The spatial dependence of the DOS often arises from an impurity scattering of the conduction electrons.  In this case the incident and scattered quasiparticle waves interfere and give rise to Friedel-like oscillations in the local density of states (LDOS). Such quasiparticle interference (QPI) effects have first been observed by STS experiments on normal state metal surfaces \cite{Crommie1993,Sprunger1997,Petersen1998}. A convenient way to extract the dominating scattering vectors $\bf q$ from a spatially resolved image of the QPI pattern is by analysis of its Fourier transformed image \cite{Sprunger1997,Petersen1998}.
This so-called \textit{spectroscopic-imaging scanning tunneling microscopy} (SI-STM) has proven to be an extremely powerful technique to investigate unconventional superconductors because the scattering rate between quasiparticle states with momenta ${\bf k}$ and ${\bf k'}$ is proportional to coherence factors $(u_{\bf k}u_{\bf k'}^*\mp v_{\bf k}v_{\bf k'}^*)$, where the $\mp$ sign is determined by the nature of the underlying scattering mechanism. The coherence factors are sensitive to the phase of the superconducting order parameter via the Bogoliubov coefficients $u_{\bf k}$ and $v_{\bf k}$ which fulfill the relation $ v_{\bf k} / u_{\bf k}=(E_{\bf k}-\xi_{\bf k}) / \Delta_{\bf k}^*$
with the quasiparticle energy $E_{\bf k}=\pm (\xi^2_{\bf k}+|\Delta_{\bf k}|^2)^{1/2}$ \cite{Bardeen1957,Tinkham}. Thus, through  the coherence factors the QPI pattern is decisively influenced by the nature of superconductivity, in particular the symmetry of the superconducting gap. Prominent examples are recent studies on the QPI of cuprate  high-temperature superconductors \cite{Hoffman2002,Hanaguri2007,Hanaguri2009} and, more recently, also of iron-based superconductors \cite{Hanaguri2010}.

LiFeAs is a  \textit{stoichiometric} superconductor which exhibits clean, charge neutral cleaved surfaces with a bulk-like electronic structure at the surface \cite{Lankau2010}. Its critical temperature is $T_c\sim18$~K \cite{Tapp2008} and currently available single crystals are of high quality \cite{Morozov2010}. With these properties LiFeAs emerges as a prime material to probe the superconducting gap characteristics using SI-STM. Our studies reveal that the QPI is governed by a van-Hove singularity at the center of the Brillouin zone where the scattering between hole-like bands dominates. Moreover, the comparison of the experimental data with QPI calculations suggests incompatibility with elementary singlet pairing symmetries ($s_\pm$,  $s_{++}$ or $d$-wave) but yields optimal agreement if an elementary $p$-wave order parameter is assumed.

We have grown high quality single crystals of LiFeAs by a self-flux technique \cite{Morozov2010}. The SI-STM measurements have been performed in a home-built apparatus at $T\approx 5.8$~K in cryogenic vacuum \cite{epaps}.

\begin{figure}[t]
\centering
\includegraphics[width=\columnwidth]{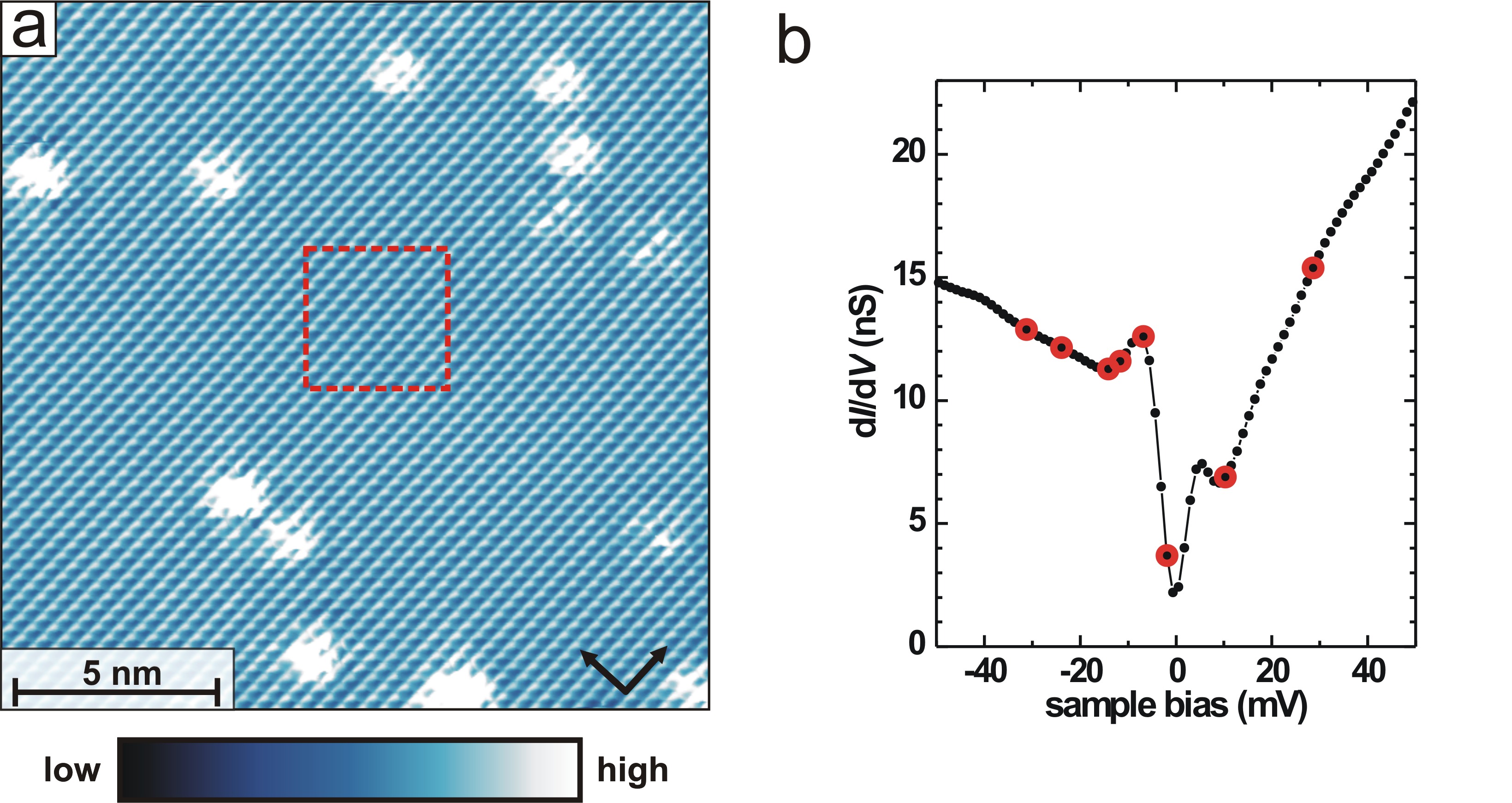}
\caption{(a) Surface topography of LiFeAs measured in constant current mode ($I=600~$pA, $V_{\mathrm{bias}}=-50$~mV) taken at $T\approx5.8$~K. Black arrows indicate the direction of the lattice constants \cite{Tapp2008} with $a=3.7914~${\AA}. (b) Spatially averaged tunneling spectrum taken in the square area (dashed lines) in (a). The spectrum exhibits a gap with $2\Delta\sim10$~mV, which (taking thermal broadening into account) is consistent with low-temperature  ($\sim 500$~mK) tunneling data \cite{Takagi2011} of the superconducting gap of LiFeAs. Representative energy values for further QPI analysis in Fig.~\ref{qpi} are marked by red circles.\label{topo}}
\end{figure}

Figure~\ref{topo}a shows a representative surface topography of LiFeAs after cleaving in cryogenic vacuum at $\sim5.8$~K, registered in constant-current mode at the bias voltage $V_{\mathrm{bias}}=-50$~mV, evidencing the pristine nature of the Li-terminated surface. The field of view exhibits a highly periodic atomic corrugation which corresponds to about 2000 atoms in the topmost layer. 14 defects are present in the field of view and appear as valleys superimposed on the atomic corrugation. This yields a defect concentration $<1\%$ and highlights the high purity of our samples. 

\begin{figure}[t]
\centering
\includegraphics[width=1\columnwidth]{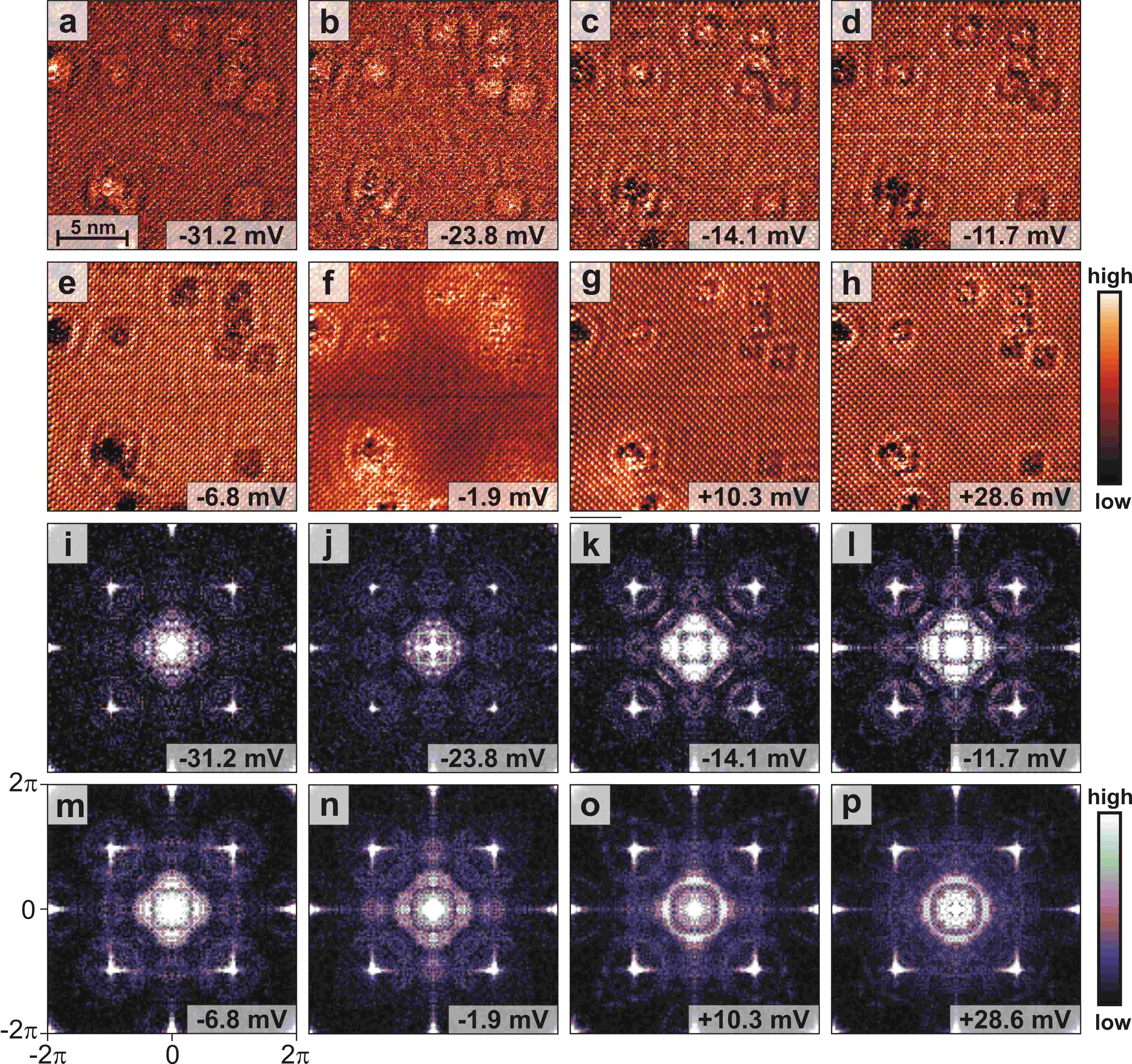}
\caption{(a-h) SI-STM maps of the region shown in Fig.~\ref{topo} at selected representative bias voltages. (i-p) Fourier transformed images of the maps shown in a-h. Bright spots at $(\pm\pi,\pm\pi)$ and at higher $\bf q$ result from the atomic corrugation in the real space images. \label{qpi}}
\end{figure}

In order to resolve the QPI we have measured the spatial- and energy resolved differential tunneling conductance, $dI/dV({\bf r}, V_{\mathrm{bias}})$ in the range $E=\rm -50~meV\dots 50~meV$ in full spectroscopic mode. A spatially averaged $dI/dV$ spectrum, taken on a defect-free area clearly reveals a superconducting gap (Fig.~\ref{topo}b). SI-STM maps at representative energies (Fig.~\ref{qpi}a-h \cite{epaps}) reveal very clear QPI patterns which are most pronounced in the vicinity of the coherence peaks at negative energy ($V_{\mathrm{bias}}\gtrsim -20$~mV). In this energy range, the QPI is clearly not only visible as relatively strong modulations close to the defects but also appears as clear wave-like modulations (with a wavelength of a few lattice spacings) in the relatively large defect-free area in the center of the field of view. QPI patterns are also discernible at positive energy, but compared to the pronounced modulations at negative values, the amplitude of the modulations decay more rapidly when moving away from a defect.

Figure~\ref{qpi}i-p displays the Fourier transformed images of the shown conductance maps. The most salient feature is a bright structure distributed around ${\bf q}=(0,0)$. In similarity to the observed real-space modulations this feature is particularly pronounced at energies $V_{\mathrm{bias}}\approx[-20~\rm mV, 0]$ where it attains a squarish shape with the corners pointing along the $(q_x,0)$ and $(0,q_y)$ directions. 
Upon increasing $V_{\mathrm{bias}}$ to positive values, the intensity at the square corners increasingly fades and for $V_{\mathrm{bias}}>10$~mV the squarish shape changes to an almost round structure which remains in that shape up to 50~mV.
The Fourier transformed images also reveal further well resolved structures with significantly lower intensity centered around $(\pi/2,\pi/2)$, $(\pi,0)$ and $(\pi,\pi)$ which again are most pronounced between -20~mV and the Fermi level. At more negative bias, these finer structures fade, while at positive bias voltage they develop into a rather featureless diffuse background.

\begin{figure*}[t]
\centering
\includegraphics[width=\textwidth]{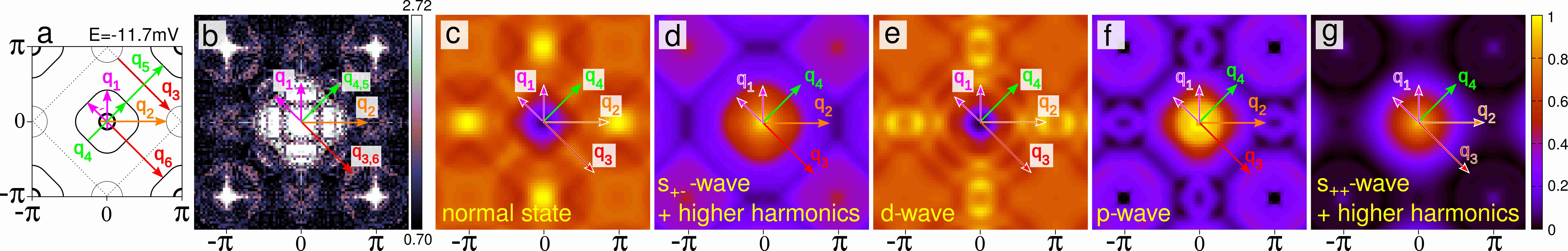}
\caption{(a) Simplified CEC \cite{Borisenko2010} at $E=-11.7$~meV in the periodic-zone scheme of the Brillouin Zone (BZ), where the first BZ (referring to the unit cell with two Fe atoms) is indicated by the dashed lines. However, the used coordinates in reciprocal space refer to the unit cell with one Fe atom, in order stay consistent with the theoretical work in Ref.~\onlinecite{Brydon2011}. This choice of reciprocal coordinates leads to Bragg-intensity at  $(\pm\pi,\pm\pi)$ instead of $(\pm2\pi,0)$ (and $(0,\pm2\pi)$) as one would expect for a two-Fe unit cell. {The two pockets around $(0,0)$ represent hole-like CEC while the pockets at the zone boundary are electron-like.}  ${\bf q}_{1,2}$ represent scattering processes which connect states on the small hole-like CEC and on other CEC, ${\bf q}_{3}$ and ${\bf q}_{4}$ represent scattering between the electron-like and  within the large hole-like CEC, respectively. ${\bf q}_{5,6}$ represent umklapp processes. Note that each scattering process ${\bf q}_{1,\dots,6}$ is described by a set of scattering vectors as is illustrated for ${\bf q}_{1}$ (dashed and solid arrows).
(b) Measured Fourier transformed image at the same energy {(the same as Figure~2l)} with ${\bf q}_{1,\dots,6}$ superimposed. The most salient QPI features around $(0,0)$ and $(\pi,0)$ match well with ${\bf q}_{1}$ and ${\bf q}_{2}$ (see text). The further observed but less prominent QPI intensities around $(\pi,\pi)$ and at $(\pi/2,\pi/2)$ are well described by ${\bf q}_3$ and ${\bf q}_4$, respectively. The umklapp scattering vectors ${\bf q}_5$ and ${\bf q}_6$ might also be of relevance here. (c-g) Calculated QPI in $\bf q$ space assuming the normal state and a superconducting order parameter with $s_\pm$-, $d$-, $p$-, and $s_{++}$ symmetry.\label{anaqpi}}
\end{figure*}

Already without considering a full calculation of the QPI in LiFeAs we can deduce important qualitative information from the data. In fact, the observed Fourier transformed images possess an eye-catching similarity to the Fermi surface of LiFeAs \cite{Borisenko2010,Brydon2011}, which immediately implies that the QPI is dominated by a set of scattering vectors which emerge from \textit{one} relatively small, high DOS region in the Brillouin zone. 
This is illustrated in more detail in Figure~\ref{anaqpi}a and b where we compare the constant energy contour (CEC) of LiFeAs at $E=-11.7$~mV \cite{Borisenko2010} with the observed QPI intensities in the Fourier transformed image. Most prominent is that the observed central squarish structure in Figure~\ref{anaqpi}b appears like a somewhat enlarged smeared replica of the large hole-like CEC around $(0,0)$ \footnote{The experimentally observed strong intensity at ${\bf q}\sim (0,0)$ reflects a constant background in the SI-STM maps which is not taken into account in our calculations. Thus there is no enhanced intensity in the calculated Fourier transformed images at $(0,0)$.}. This observation can directly be understood as stemming from scattering processes (${\bf q}_1$) connecting the very small and the larger squarish-shaped hole-like CEC around $(0,0)$. Furthermore, 
 the much weaker structure at ${\bf q}=(\pi,0)$ in Figure~3b, which resembles a smeared version of the electron-like CEC near ${\bf k}=(\pi,0)$, apparently results from scattering processes (${\bf q}_2$) connecting  this CEC with again the small hole-like CEC. Note that due to the presence of a van-Hove singularity \cite{Borisenko2010} at ${\bf k} = (0,0)$ the DOS for momentum vectors on the small hole-like CEC is enhanced significantly. Thus, scattering processes combining those momenta with {\bf k}-points close to the other CEC are dominant and give rise to a Fourier transformed QPI image which essentially represents a 'map' of all CEC at a given energy.  

Having assigned the main features of the measured QPI patterns to the normal state electronic structure, the next step is to establish how in the superconducting state the DOS is redistributed by the opening of the superconducting gap. One expects that in the superconducting state the DOS at energy values close to the gap value is further boosted in comparison to the normal state since the quasiparticle dispersion $E_{\bf k}=\pm (\xi^2_{\bf k}+|\Delta_{\bf k}|^2)^{1/2}$ is rather flat in the vicinity of the Fermi surface. Furthermore, depending on the gap function $\Delta_{\bf k}$ particular scattering channels are suppressed while others are enhanced according to the coherence factors. Consequently, the QPI measured at energies $|E|$ close to the averaged gap value will be enhanced significantly as compared to the normal state. This intensity redistribution contains detailed information about the structure of the superconducting order parameter.

For a quantitative analysis of the experimental data, we calculated the QPI in the superconducting state using an appropriate BCS model for LiFeAs which can describe three cases of elementary singlet pairing ($s_{++}$, $s_\pm$, and $d$-wave) as well as a $p$-wave triplet pairing scenario. {To ensure that our calculations are based on unbiased experimental findings, the electronic band structure of LiFeAs was modeled using tight-binding fits derived from recent ARPES experiments \cite{Borisenko2010}. We note that our band structure model obtained in this way is, at least for the one-particle states to be considered here, consistent with an alternative three band orbital model \cite{Brydon2011} for LiFeAs.}  Following Refs.~\onlinecite{Maltseva2009} and \onlinecite{Sykora2010}, we numerically evaluated the LDOS in momentum space. The scattering potential of the impurities was modeled by a conventional non-magnetic local Coulomb potential \cite{epaps}.

In Fig.~\ref{anaqpi}c-g we compare the results of our calculations for the energy $E=-11.7$~mV. The calculated QPI pattern of the normal state shown in Figure~\ref{anaqpi}c exhibits an extended low-intensity region around $(0,0)$, in stark contrast to the experimental data (Fig.~\ref{anaqpi}b). 
The situation drastically changes if the superconducting state is taken into account in the QPI calculations. Figure~\ref{anaqpi}d-f shows the calculated QPI for four different types of the superconducting order parameter.
The main feature, which now can be observed for three of the four superconducting cases is a wide high-intensity region centered at $(0,0)$. This qualitative change clearly corroborates that the intense central  structure observed in the experiment is characteristic of the superconducting state and a comparison of Figure~\ref{anaqpi}e with the experimental result clearly allows us to exclude a {clean} $d$-wave pairing in LiFeAs.  
Furthermore, from the location of the high intensity region we conclude that in the superconducting state the dominating scattering processes are those which connect states on the hole-like bands. 
The prevalence of these scattering processes is naturally explained by the presence  of the van Hove singularity in the center of the Brillouin zone and is one of our major findings. It implies that nesting-enhanced scattering, which is predominant and crucial for $s_\pm$-superconductivity in other iron pnictides \cite{Mazin2008,Mazin2010,Johnston2010,Hanaguri2010}, is unimportant in LiFeAs.

\begin{figure}[t]
\centering
\includegraphics[width=1\columnwidth]{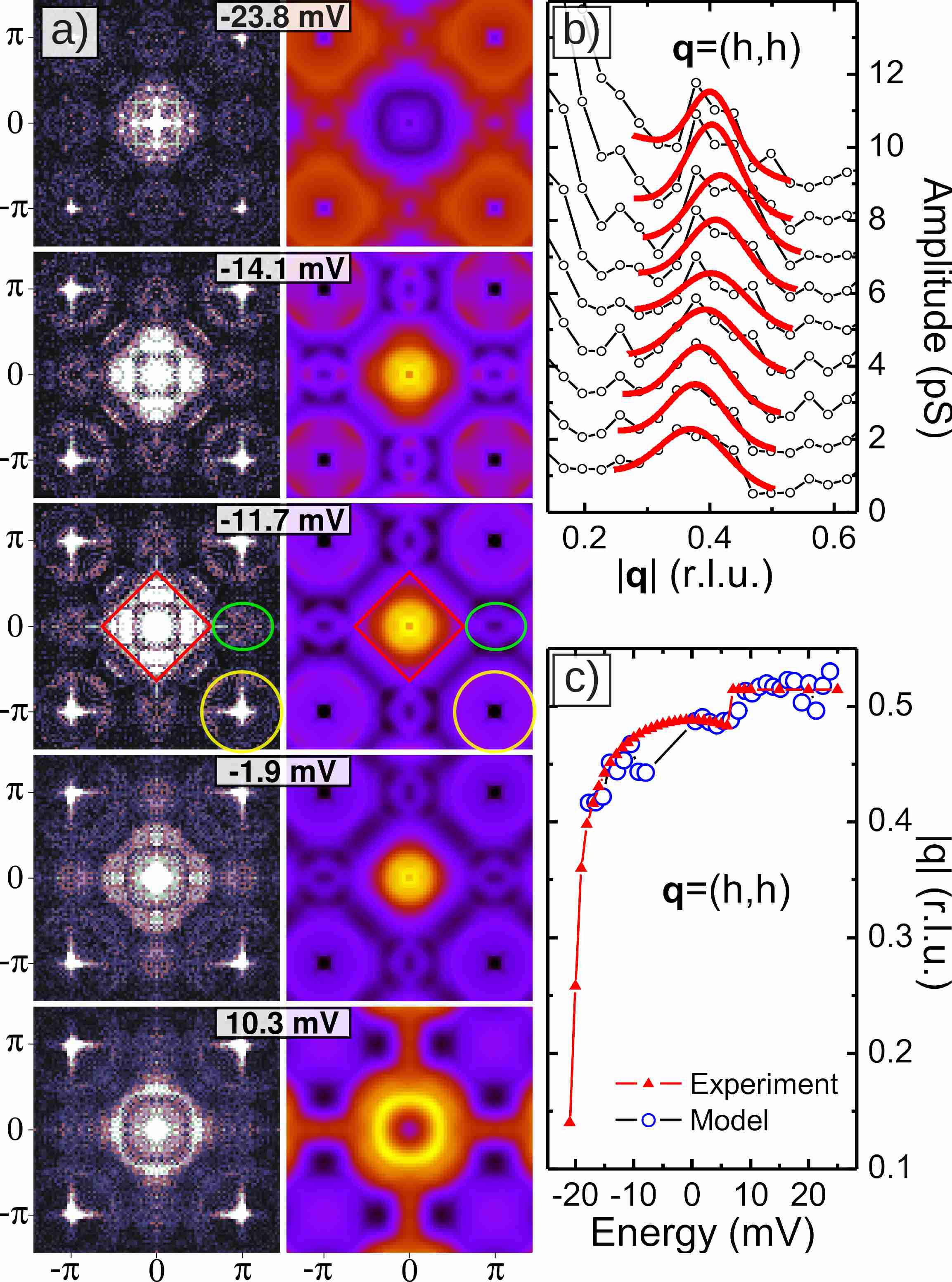}
\caption{(a) Comparison of the measured QPI pattern and its calculated counterpart based on a chiral $p$-wave superconducting order parameter vor various energies. The clearly observable dispersion unambiguously demonstrates that the observed structures result from QPI and not from non-dispersing modifications of the DOS around a defect. b) Representative experimental scans along ${\bf q} =(h,h)$ with Gaussian fits which yield the edge position of the measured central structure along this direction. c) Size of the measured and calculated central structure along ${\bf q} =(h,h)$.\label{dispq1}}
\end{figure}

A closer examination of the individual QPI patterns brings to light a very strong dependency of the pattern on the assumed pairing symmetry. This allows a direct comparison with the experimental result. Interestingly, we find a striking agreement between the experimentally observed QPI and the calculated image of Fig.~\ref{anaqpi}f, which is based on a triplet paired superconducting state. In this case we considered a gap function which has a chiral $p$-wave symmetry to be consistent with the experimental observation of lacking gap nodes \cite{Borisenko2010}. The agreement is obviously much less pronounced for $s_\pm$-, $s_{++}$ and $d$-wave singlet pairing. The main bright feature around $(0,0)$ is absent for the $d$-wave or rather circular for the $s_{++}$-wave cases where we have included higher harmonics in order to account for the experimentally observed variation of the gap size along the Fermi surface.  One could argue, that the intensity distribution around $(0,0)$ found for the assumed $s_\pm$-pairing somewhat resembles the experimental ﬁndings. However, the well resolved structures which are described by the scattering vectors ${\bf q}_2$ and ${\bf q}_{3,6}$ which are marked by green and yellow lines in Fig.~\ref{dispq1}(a) are consistent with the experimental result only for the $p$-wave case. Thus, among the three considered order parameters the best agreement between experimental data and calculated QPI images is apparently obtained for a $p$-wave order parameter.
 
To show that this surprising similarity holds also at other energies, we present in Fig.~\ref{dispq1} a comparison between experimental results and calculated $p$-wave based QPI for another four energy values. We find an excellent agreement for the energy evolution of the three most intensive QPI structures (framed in Fig.~\ref{dispq1}(a)) which are related to the scattering vectors ${\bf q}_1$, ${\bf q}_2$ and ${\bf q}_{3,6}$ (compare Fig.~\ref{anaqpi}(a)). A closer inspection of the boundary of the central structure which is shown in Fig.~\ref{dispq1}(b) reveals a clear dispersive behavior of the high-intensity region within some finite energy range. As it is presented in Fig.~\ref{dispq1}(c) this energy dependence is confirmed very well within our model calculations.

Our main observation that the quasiparticle scattering is dominated by the van-Hove singularity at the center of the Brillouin zone fits in very well in recent experimental and theoretical findings. Angle resolved photo emission spectroscopy (ARPES) data provide clear-cut evidence for the absence of strong nesting in LiFeAs \cite{Borisenko2010}. From these experiments an $s_\pm$ pairing scenario, which is believed to be nesting-driven by antiferromagnetic fluctuations at ${\bf q}=(\pi,0)$ \cite{Mazin2008}, is not expected.
In fact, a recent theoretical analysis based on a realistic band structure model with poor nesting yields a dominating role of the small hole-like Fermi surface for the scattering in LiFeAs, giving rise to ferromagnetic fluctuations which drive an instability towards triplet superconductivity \cite{Brydon2011}. It seems rather encouraging that our QPI calculations which are based on a quite simple model and very elementary assumptions (potential scattering, i.e., no specific presumptions for the scattering centers as well as elementary order parameters) are consistent with this finding. It is, however, thinkable that a more complex order parameter (such as $s+id$ symmetry) and/or a specific character of the impurities would yield a similar convincing agreement. Thus, further experimental and theoretical studies are indispensable for clarifying the nature of superconductivity in LiFeAs.

\section*{Acknowledgements}
The authors thank M. Vojta, C. Timm, D. Morr, S. Aswartham, C. Nacke, I. Morozov, V. Zabolotnyy, S. Borisenko and H. Takagi for valuable
discussions and comments. Furthermore, we thank S. Pichl, J. Werner, M. Deutschmann for sample characterization and  V. Zabolotnyy, S. Borisenko for providing band structure data of LiFeAs. This work has been supported by the Deutsche Forschungsgemeinschaft through the Priority Programme SPP1458 (Grants No. BE1749/13 and No. GR3330/2), and through the Emmy Noether Programme (Grants No. WU595/3-1 [S. W.] and No. DA1235/1-1 [M. D.]).


\newpage

\section*{Experimental details}
High quality single crystals of LiFeAs with a critical temperature $T_c\approx 18$~K have been grown by a self-flux technique \cite{Morozov2010}. The chemical and physical properties of the crystals were studied by means of energy-dispersive X-ray (EDX) analysis, inductively coupled plasma mass spectroscopy (ICP MS), single crystal diffractometry, nuclear quadrupole resonance (NQR), magnetization, specific heat, electrical transport and angle resolved photo emission spectroscopy (ARPES) \cite{Morozov2010,Borisenko2010}.

The crystals were inserted into a home-built low-temperature scanning tunneling microscope and cooled down to $T\approx 5.8$~K where the samples were cleaved in cryogenic vacuum. Subsequently scanning tunneling microscopy (STM) and -spectroscopy (STS) measurements were performed. After cleavage the sample surface exhibited extended areas with atomic corrugation. A rather high density of further objects, presumably left-over atoms from the cleaving process, were present in the first topographic measurement after cleavage. A part of these objects were highly mobile and were 'pushed' out of the investigated area by extensive scanning. This resulted in the topographic data shown in Fig.~1, which displays an atomically flat surface containing the \textit{immobile} impurities which give rise to the quasiparticle interference.

The SI-STM maps shown in Fig.~2 have been acquired in full spectroscopic mode, i.e., by measuring STS at each of the $256\times256$ pixel by stabilizing the tip with feedback loop on at a setpoint  of $V_{\mathrm{bias}}=-50$~mV and $I=600$~pA followed by subsequently ramping $V_{\mathrm{bias}}$ to $+50$~mV with the feedback loop switched off. During ramping the voltage we recorded $I(V_{\mathrm{bias}})$ and $dI/dV(V_{\mathrm{bias}})$ where the latter was recorded using a lock-in amplifier with a modulation voltage $V_{\mathrm{mod}}=1.2$~mV (RMS) and a modulation frequency $f_{\mathrm{mod}}=3.333$~kHz.

The Fourier transforms in Fig.~2i-p were calculated from the SI-STM maps in Fig.2~a-h after filtering the data using a Hamming window in order to suppress edge effects. Since the resulted Fourier transformed images were slightly distorted due to piezo relaxation effects, the images were first linearly transformed in order to restore the tetragonal lattice. Afterwards each Fourier transformed image was symmetrized with respect to the high symmetry in-plane directions.

\newpage
\section*{Model of the Quasiparticle Interference} 
Following the ideas of Refs.~\onlinecite{Maltseva2009} and \onlinecite{Sykora2010} we 
developed a theoretical model to calculate the tunneling conductance which is measured by an STM 
experiment. Within a simplified model of the tunneling process the differential tunneling 
conductance $dI / dV (\mathbf{r},V)$ at a location $\mathbf{r}$ and
Voltage $V$ is proportional to the single-particle density 
of states at energy $\omega$. The spatial dependence of the tunneling-matrix element, 
which includes contributions of the sample wave function around the 
tip, has been neglected within the calculations. To study the quasiparticle 
interference we calculated the
fluctuations of the density of states (LDOS) arising from an electron scattering 
off impurities which are distributed in the sample.

The fluctuations of the LDOS are determined by the Green's function of a superconductor in the presence of impurities. Assuming the electron field inside a superconductor can be described by a four-component vector which is usually written in Nambu notation as
\begin{equation}
\label{spinor_r}
\Psi^\dag(\mathbf{r},\tau) =
\left(
\psi_{\uparrow}^\dag ({\bf r},\tau ), 
\psi_{\downarrow}^\dag({\bf r},\tau ),
\psi_{\downarrow}({\bf r},\tau ), 
-\psi_{\uparrow}({\bf r},\tau )
\right) ,
\end{equation}
where $\mathbf{r}$ denotes the position vector
and $\tau$ the imaginary time,  
the matrix Green's function is defined as the time-ordered average,
\begin{equation}
\label{2}
\hat{G}_{\alpha\beta}(\mathbf{r}',\mathbf{r};\tau) = -\langle T_\tau \Psi_{\alpha}(\mathbf{r}',\tau)
\Psi_{\beta}^\dag(\mathbf{r},0)\rangle.
\end{equation}
As it is usual for an electronic system in the presence of impurities we calculate the Green's function 
\ref{2} using the t-matrix method. For the dominant effect the scattering potential is taken into 
account in Born approximation. Within this approximation the Fourier transformed Green's function can be written as
\begin{equation}
\label{G_fluct}
\hat{G}(\mathbf{k},\mathbf{k}',\omega) = \hat{G}_0(\mathbf{k},\omega) + \hat{G}_0(\mathbf{k},\omega)
\hat{U}(\mathbf{k},\mathbf{k}',\omega) \hat{G}_0(\mathbf{k}',\omega),
\end{equation}
where $\hat{U}(\mathbf{k},\mathbf{k}')$ is the scattering potential of a single impurity and 
$\hat{G}_0(\mathbf{k},\omega)$ is the bare Green's function of the BCS superconductor without 
impurities. The LDOS is determined by the analytic continuation $\hat{G}(\mathbf{r}',\mathbf{r};i\omega_n) 
\rightarrow \hat{G}(\mathbf{r}',\mathbf{r};z)$ of the Matsubara Green's function
$\hat{G}(\mathbf{r}',\mathbf{r};i\omega_n) = \int_0^\beta \hat{G}(\mathbf{r}',\mathbf{r};\tau) 
e^{i(2n + 1) \pi T \tau} d\tau $,
\begin{equation}
\label{LDOS_r}
\rho(\mathbf{r},\omega) = \frac{1}{\pi} \mbox{Im} \mbox{Tr} \frac{\hat{1} + \hat{\tau}_3}{2} 
\left[\hat{G}(\mathbf{r},\mathbf{r}; \omega - i \delta)\right],
\end{equation}
where $\hat{1}$ is the $4\times 4$ unit matrix and $\hat{\tau}_3$ is
the isospin matrix, 
\begin{equation}
\label{tau_3}
\hat{\tau}_3 = \left(
\begin{array}{cr}
\underline{1} & \underline{0} \\
\underline{0} & -\underline{1} 
\end{array}
\right), \nonumber
\end{equation}
where an  underscore denotes a two dimensional matrix. Taking the Eqs. \ref{G_fluct} and 
\ref{LDOS_r} we evaluated the LDOS using a BCS model for a triplet superconductor which is based on a realistic three band model for LiFeAs taken from Ref.~\onlinecite{Brydon2011}. Note that the band structure of the used model is consistent with recent ARPES experiments. For the superconducting order parameter we assume a chiral $p$-wave symmetry with the following d-vector,
\begin{equation}
{\bf d}({\bf k}) = \Delta [\sin (k_x) \pm i \sin (k_y)] \vec{e}_z, \nonumber
\end{equation}
where $\Delta$ is chosen such that the corresponding gap size in the calculated conductance $dI/dV(\omega)$ coincides with the experimental one shown in Fig.~\ref{dispq1}. Note that within our simplified model the conductance is assumed to be proportional to the local density of states which is given by Eq.~\ref{LDOS_r}.  

By assuming that the scattering potential of the impurities is screened at length scales comparable to the lattice spacing we model the impurity scattering by a local Coulomb potential which can be described by a matrix of the form $\hat{U} = U \hat{\tau}_3$. Note that for a local potential the momentum dependence of  $\hat{U}$ can be omitted. Furthermore, for simplicity, we assumed that the scattering is equal for all electron bands.\\\\

\end{document}